\documentclass [11pt]{article}

\usepackage{doublespace}
\usepackage{epsfig}
\usepackage{amssymb}
\usepackage{amsthm}
\usepackage{verbatim}
\usepackage{delarray}
\usepackage{graphics}
\usepackage{epic}

\newcommand{\ket}[1]{{|#1\rangle}}
\newcommand{\bra}[1]{{\langle#1|}}
\newcommand{\braket}[2]{{\langle#1|#2\rangle}}

\newcommand{\bl}{\left(}
\newcommand{\br}{\right)}

\newcommand{\tr}{\mbox{Tr}}

\newcommand{\deft}{\ {\stackrel{\triangle}{=}} \ }

\newcommand{\hpi}{\widehat{\Pi}}
\newcommand{\hx}{\widehat{X}}

\newcommand{\HH}{{\mathcal{H}}}

\newcommand{\B}{{\mathcal{B}}}

\newcommand{\I}{{\mathcal{I}}}
\newcommand{\C}{{\mathbb{C}}}

\newcommand{\G}{{\mathcal{G}}}
\newcommand{\Q}{{\mathcal{Q}}}
\newcommand{\SSS}{{\mathcal{S}}}

\newcommand{\ie}{{\em i.e., }}
\newcommand{\eg}{{\em e.g., }}

\newcommand{\etal}{\emph{et al.\ }}

\newtheorem{theorem}{Theorem}

\title{Optimal Detection of Symmetric Mixed Quantum States} 
\author{Yonina
C. Eldar\footnote{The author was with the Research Laboratory of
Electronics, Massachusetts Institute of Technology, Cambridge, MA and
is now with the Technion, Israel Institute of Technology, Haifa
32000, Israel. E-mail: yonina@ee.technion.ac.il.}, Alexandre
Megretski\footnote{Laboratory for Information and Decision Systems,
Massachusetts Institute of Technology, Cambridge, MA 02139. E-mail:
ameg@mit.edu.},  and George C. Verghese\footnote{
Laboratory for Electromagnetic and Electronic Systems,
Massachusetts Institute of Technology,
Cambridge, MA 02139. E-mail: verghese@mit.edu.}
}

\date{\today}
 
\begin{document} 

\maketitle

\begin{abstract}
We develop a sufficient condition for the least-squares measurement
(LSM), or the square-root measurement, to minimize the probability of
a detection error when distinguishing between a collection of mixed
quantum states.   Using this condition we derive the optimal  
measurement for state sets with a broad class of symmetries.

We first consider geometrically uniform (GU) state sets with a
possibly nonabelian generating group, and show that if the generator
satisfies a certain constraint, then the LSM is  optimal. In
particular, for pure-state GU ensembles the LSM is shown to be optimal.
For arbitrary GU state sets we show that the optimal measurement
operators are GU with generator that can be computed very efficiently
in polynomial time, within any
desired accuracy. 

We then consider compound GU (CGU) state sets which consist of subsets that
are GU.  When the generators satisfy a certain constraint, the
LSM is again optimal.  For arbitrary CGU state sets the optimal
measurement operators are shown to be CGU with generators that can be
computed efficiently in polynomial time.

\end{abstract}

%%%%%%%%%%%%%%%%%%%%%%%%%%
\section{Introduction}
\label{sec:intro}
%%%%%%%%%%%%%%%%%%%%%%%%%%

In a quantum detection problem a transmitter conveys
classical information to 
a receiver using a quantum-mechanical channel. Each message
corresponds to a preparation of
the quantum channel in an associated quantum
state represented by a density operator,  
drawn from a collection of known states. 
To detect the information, the receiver  subjects the channel to a
quantum measurement. 
Our problem  is to construct a
measurement that minimizes the
probability of a detection error.

We consider a quantum state ensemble consisting of $m$ 
density operators $\{\rho_i, 1
\le i \le m\}$ on an $n$-dimensional complex Hilbert space $\HH$, with 
prior probabilities $\{p_i>0, 1 \le i \le m\}$.
A density operator $\rho$ is a 
positive
semidefinite (PSD)
 Hermitian operator with $\tr(\rho)=1$; we write $\rho \geq 0$ to
indicate $\rho$ is PSD. A mixed state ensemble is one in which at
least one of the density operators $\rho_i$ has rank larger than one.
A pure-state ensemble
is one in which each density operator $\rho_i$ is a rank-one projector
$\ket{\phi_i}\bra{\phi_i}$, where the vectors $\ket{\phi_i}$, though
evidently normalized to unit length, are
 not necessarily orthogonal.

For our {\em measurement} we  consider general
positive operator-valued measures \cite{H76,P90}.
Necessary and sufficient conditions for an optimum measurement
minimizing the probability of a detection error have been derived
\cite{H73,YKL75,EMV02}.  
However, in general,
obtaining a closed-form analytical expression for the optimal
measurement directly from these conditions is 
a difficult and unsolved problem. Iterative algorithms minimizing the
probability of a detection error have been proposed in \cite{H82,EMV02}.

There are some particular cases in which the solution to the quantum
detection problem is known explicitly \cite{H76,CBH89,OBH96,BKMO97,EF01}.
Ban \etal \cite{BKMO97} derive the solution for  a pure-state ensemble 
consisting of density operators $\rho_i=\ket{\phi_i}\bra{\phi_i}$
where the 
vectors 
$\ket{\phi_i}$ form a cyclic set, \ie the vectors  are generated
by a cyclic group of unitary matrices using a single generating vector.
 The optimal measurement coincides with 
the least-squares measurement (LSM) \cite{EF01}, also known as the 
square-root measurement \cite{HW94,H96}. 
Eldar and Forney \cite{EF01} derive the optimal measurement for a
pure-state ensemble in which the vectors $\ket{\phi_i}$  
have a strong symmetry property called geometric uniformity.
In this case the vectors $\ket{\phi_i}$ are
defined over a finite abelian group of unitary matrices and generated by a
single generating vector; the optimal
measurement again coincides with the LSM. Note, that a cyclic state
set is a special case of a geometrically
uniform state set.
 
The LSM  has many desirable properties
\cite{EF01,HW94,H96,BK00,BKMO97,CP02,H98} 
and has therefore been proposed
as a detection measurement in 
many settings (see \eg \cite{SKIH98,SUIH98,KOSH99}).  
In Section~\ref{sec:lsm} we derive a sufficient condition on the
density operators for the LSM
to minimize the probability of a detection error.
For rank-one ensembles we show that
the LSM minimizes the probability of a detection error if the
probability of correctly detecting each of the states using the LSM is
the same, regardless of the state transmitted.

In Section~\ref{sec:gu} we consider geometrically uniform (GU) state
sets defined over a finite group of unitary matrices. In contrast to
\cite{EF01}, the GU state sets we
consider are not constrained to be rank-one state sets but rather can
be mixed state sets, and the unitary
group is not constrained to be abelian.
We obtain a convenient
characterization of the LSM and show that the LSM operators have the
same symmetries as the original state set. We then show that for such
GU state sets the probability of correctly detecting each of the states
using the LSM is the same, so that for rank-one ensembles, the 
LSM minimizes the probability of a detection error. For an arbitrary
GU ensemble, the optimal measurement operators are shown to be  GU
with the same 
generating 
group, and can be computed very efficiently in polynomial
time. Furthermore, under a certain constraint on the generators,
the LSM again minimizes the
probability of a detection error.

In Section~\ref{sec:cgu} we consider the case in which the state set
is generated by a group of unitary matrices using {\em
multiple} generators. Such a collection of
states is referred to as a {\em compound GU (CGU)} state set
\cite{EB01}.  We obtain a convenient
characterization of the LSM for CGU state sets, and show that the LSM
vectors are themselves CGU.  
When the probability of correctly detecting each of the {\em generators}
using the LSM is the same, we show that the probability of correctly
detecting each of the states using the LSM is the same. Therefore, for
rank-one CGU ensembles with this property, the LSM minimizes
the probability of a detection error. 
An interesting class of CGU state sets results when the
set of generating vectors is itself GU, which we refer to as {\em CGU
state sets with GU generators}. In the case in which the generating
vectors are GU 
and generated by a group that commutes up to a phase factor with the
CGU group, we show that the LSM vectors are also CGU with GU generators
so that they are generated by a {\em
single} generating vector. 
For such state sets, the probability of correctly detecting each of
the states 
using the LSM is the same, so that for rank-one ensembles, the 
LSM minimizes the probability of a detection error.
Finally we show that for arbitrary CGU state sets, the
measurement operators 
minimizing the probability of a detection error are also CGU, and we
propose an efficient algorithm for computing the optimal generators.

Before proceeding to the detailed development, in the next section we
present our problem and summarize results from \cite{EMV02}
pertaining to the conditions on the optimal measurement operators.

%%%%%%%%%%%%%%%%%%%%%%%%%%%%%%%%%%%%%%%%%%%%%%%
\section{Optimal Detection of Quantum States} 
\label{sec:qd} 
%%%%%%%%%%%%%%%%%%%%%%%%%%%%%%%%%%%%%%%%%%%%%%%

Assume that a quantum channel is prepared in a 
quantum state drawn from a collection of given states represented by
density operators $\{
\rho_i,1 \leq i \leq m \}$ in  an $n$-dimensional
complex Hilbert 
space $\HH$.
We assume without loss of generality that the eigenvectors of
$\rho_i,1 \leq i \leq m$, collectively 
span\footnote{Otherwise we can transform the problem to a problem
equivalent to the one considered in this paper by reformulating the
problem on the subspace spanned by the eigenvectors of  $\{\rho_i,1
\leq i \leq m\}$. } $\HH$ so that $m \geq n$. 
Since $\rho_i$ is Hermitian and PSD, we can express $\rho_i$ as 
$\rho_i=\phi_i\phi_i^*$ for some matrix $\phi_i$, \eg via the Cholesky or
eigendecomposition of $\rho_i$ \cite{GV96}. We refer to $\phi_i$
as a {\em factor} of $\rho_i$. Note that the choice of $\phi_i$ is
not unique; if $\phi_i$ is a factor of $\rho_i$, then any matrix of
the form $\phi_i'=\phi_iQ_i$ where $Q_i$ is an arbitrary matrix
satisfying $Q_iQ_i^*=I$, is also a factor of $\rho_i$.

At the receiver, the constructed
measurement comprises $m$ 
PSD
Hermitian measurement
operators $\{\Pi_i,1 \leq i \leq m\}$ on $\HH$ that satisfy
$\sum_{i=1}^m \Pi_i =I$,
where $I$ is the identity operator on $\HH$. 
We seek the measurement operators $\{\Pi_i,1 \leq i \leq m\}$
satisfying 
\begin{eqnarray}
\label{eq:identu}
\Pi_i & \geq & 0,\quad 1 \leq i \leq m;\nonumber \\
\sum_{i=1}^m \Pi_i & = & I,
\end{eqnarray}
that
minimize the probability of  
a detection error, or equivalently, maximize the probability of correct
detection. Given that the transmitted state is $\rho_j$, the
probability of correctly detecting the state using measurement
operators $\{\Pi_i,1 \leq i \leq m\}$ is
$\tr(\rho_j\Pi_j)$. Therefore, the probability of correct detection is
given by
\begin{equation}
\label{eq:pe}
P_d=\sum_{i=1}^mp_i\tr(\rho_i\Pi_i),
\end{equation}
where $p_i>0$ is the prior probability of $\rho_i$, with $\sum_i p_i=1$. 

It was shown in \cite{YKL75,EMV02} that a set of measurement operators
$\{\hpi_i,1 \leq i \leq m\}$
minimizes the probability of a detection error for a state set
$\{\rho_i,1 \leq i \leq m\}$ with prior probabilities $\{p_i,1 \leq i
\leq m\}$ if and only if there exists an Hermitian $\hx$ satisfying 
\begin{equation}
\label{eq:condhx}
\hx \geq p_i \rho_i,\quad 1 \leq i \leq m,
\end{equation}
such that 
\begin{equation}
\label{eq:condz}
(\hx-p_i\rho_i)\hpi_i=0,\quad 1 \leq i \leq m.
\end{equation}
The matrix $\hx$ can be determined as the solution to the problem
\begin{equation}
\label{eq:min}
\min_{X \in \B} \tr(X) 
\end{equation}
where $\B$ is
the set of Hermitian operators on $\HH$,
subject to
\begin{equation}
\label{eq:condx}
X \geq p_i\rho_i,\quad 1 \leq i \leq m.
\end{equation}

Except in some
particular cases \cite{H76,CBH89,OBH96,BKMO97,EF01}, obtaining a
closed-form analytical expression for the optimal measurement operators
directly from these necessary and sufficient conditions for optimality
is a difficult
and unsolved problem. 
Since 
(\ref{eq:min}) is a (convex) semidefinite programming
\cite{VB96,A91t,NN94} problem, 
there are very 
efficient methods for solving (\ref{eq:min}). 
In particular, the optimal matrix $\hx$ minimizing
$\tr(X)$ subject to (\ref{eq:condx}) can be computed in Matlab using
the linear matrix inequality (LMI) Toolbox. A 
convenient interface for using the LMI toolbox is the 
Matlab package\footnote{This software was created by A.~Megretski,
C-Y.~Kao, U.~J\"{o}nsson and A.~Rantzer and is available at
\texttt{http://www.mit.edu/cykao/home.html}.}  IQC$\beta$ (see
\cite{EMV02} for further details).
Once we  determine $\hx$, 
the optimal measurement operators
$\hpi_i$ can be computed using 
(\ref{eq:condz}) and  (\ref{eq:identu})
as described in \cite{EMV02}.

A suboptimal measurement that has been employed as a detection
measurement in many applications and has many desirable properties is
the LSM \cite{EF01,H98}.  Using the necessary and sufficient
conditions (\ref{eq:identu}), (\ref{eq:condhx}) and 
(\ref{eq:condz}), in Section~\ref{sec:lsm} we derive a general
condition under which the LSM is optimal, \ie minimizes the
probability of a detection error when distinguishing between possibly
mixed quantum states. In Sections~\ref{sec:gu} and
\ref{sec:cgu} we consider some special cases of mixed and pure state
sets for which 
the LSM is optimal, and derive explicit formulas for the optimal
measurement operators.

\newpage
%%%%%%%%%%%%%%%%%%%%%%%%%%%%%
\section{The LSM and the Optimal Measurement}
\label{sec:lsm}
%%%%%%%%%%%%%%%%%%%%%%%%%%

The LSM corresponding to a set of density operators
$\{\rho_i=\phi_i\phi_i^*,1 \leq i \leq m\}$ with eigenvectors that
collectively 
span $\HH$ and prior probabilities $\{p_i,1 \leq i \leq m\}$ consists
of the measurement operators $\{\Sigma_i=\mu_i\mu_i^*,1 \leq i \leq
m\}$ where 
\cite{H98,EF01}
\begin{equation}
\label{eq:glsm}
\mu_i=(\Psi\Psi^*)^{-1/2}\psi_i \deft T \psi_i,
\end{equation}
with
\begin{equation}
\label{eq:T}
T=(\Psi\Psi^*)^{-1/2}.
\end{equation}
Here $\Psi$ is the matrix of (block) columns $\psi_i=\sqrt{p_i}\phi_i$ and
$(\cdot)^{1/2}$ is the unique Hermitian square root of the
corresponding matrix. Note that since the eigenvectors of the $\{\rho_i\}$
collectively 
span $\HH$, the columns of the $\{\psi_i\}$ also together span $\HH$,
so $\Psi\Psi^*$ 
is invertible. 
From (\ref{eq:glsm}),
\begin{equation}
\sum_{i=1}^m \mu_i\mu_i^*=
(\Psi\Psi^*)^{-1/2}\bl \sum_{i=1}^m \psi_i\psi_i^* \br
(\Psi\Psi^*)^{-1/2}=
(\Psi\Psi^*)^{-1/2}\Psi\Psi^*
(\Psi\Psi^*)^{-1/2}=I,
\end{equation}
so that the LSM operators defined by (\ref{eq:glsm}) satisfy
(\ref{eq:identu}).
In the case in which the prior probabilities are all equal,
\begin{equation}
\label{eq:lsmep}
\mu_i=(\Phi\Phi^*)^{-1/2}\phi_i,
\end{equation}
where $\Phi$ is the matrix of (block) columns $\phi_i$.

Since the factors $\phi_i$ are not unique, the LSM factors $\mu_i$ are
also not unique. In particular, if $\mu_i$ are the LSM factors corresponding to
$\phi_i$, then the LSM factors corresponding to $\phi'_i=\phi_iQ_i$
with $Q_iQ_i^*=I$ are $\mu_i'=\mu_iQ_i$. Therefore, although the LSM
factors are not unique, the LSM
operators $\Sigma_i=\mu_i\mu_i^*$ are unique.

The LSM corresponding to a  pure-state ensemble $\ket{\phi_i}$ consists
of the measurement 
vectors $\ket{\mu_i}=T\ket{\psi_i}$, where
$\ket{\psi_i}=\sqrt{p_i}\ket{\phi_i}$. 
It was shown in \cite{EF01} that for rank-one ensembles the LSM
vectors $\ket{\mu_i}$ minimize  
the sum of the squared 
norms of the error vectors $\ket{e_i}=\ket{\mu_i}-\ket{\psi_i}$, so
that they are 
the  measurement vectors that satisfy (\ref{eq:identu}), and are
closest in a squared error sense to the 
weighted state vectors $\ket{\psi_i}=\sqrt{p_i}\ket{\phi_i}$. 
In the
case in which the vectors $\ket{\phi_i}$ are linearly independent so
that $n=m$, (\ref{eq:identu}) implies that the vectors 
$\ket{\mu_i}$ must be orthonormal, so that
the LSM vectors are the closest orthonormal vectors to the
vectors $\ket{\psi_i}$ in a least-squares sense, as 
illustrated in
Fig.~\ref{fig:lsm}. A similar result was obtained for the LSM factor
$\mu_i$ corresponding to a mixed-state ensemble  with
factors $\phi_i$ \cite{CP02}.
\setlength{\unitlength}{.2in}
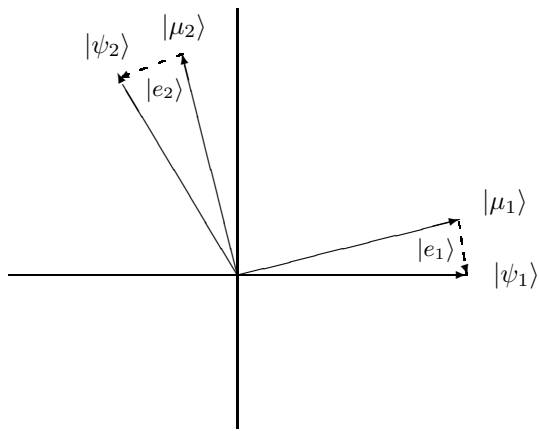
\begin{figure}[h]
\begin{center}
\begin{picture}(15,14)(0,0)
%\put(0,0){\framebox(15,14){}}
\put(7,6){\line(0,1){7}}
\put(7,6){\line(0,-1){4}}
\put(7,6){\line(-1,0){6}}
\put(7,6){\line(-3,5){3}}
\put(4.06,10.9){\vector(-1,2){.2}}
\put(3.6,12){\makebox(0,0){\small $\ket{\psi_2}$}}
\put(7,6){\vector(1,0){6}}
\put(14.3,6){\makebox(0,0){\small $\ket{\psi_1}$}}
\put(7,6){\vector(4,1){5.82}}
\put(14,7.9){\makebox(0,0){\small $\ket{\mu_1}$}}
\put(7,6){\vector(-1,4){1.45}}
\put(5.6,12.5){\makebox(0,0){\small $\ket{\mu_2}$}}

\dashline{.2}(13,6)(12.8,7.4)
\put(12.93,6.3){\vector(1,-3){.1}}
\put(12.2,6.65){\makebox(0,0){\footnotesize $\ket{e_1}$}}
\dashline{.2}(5.58,11.8)(4,11.2)
\put(4,11.2){\vector(-2,-1){.1}}
\put(5.1,10.85){\makebox(0,0){\footnotesize $\ket{e_2}$}}

\end{picture}
\caption{Example of the least-squares measurement (LSM). 
Since the vectors $\ket{\psi_i}$ are linearly independent, the LSM vectors 
$\ket{\mu_i}$ are orthonormal and minimize $\sum_i \braket{e_i}{e_i}$
where 
$\ket{e_i}=\ket{\psi_i}-\ket{\mu_i}$.}
\label{fig:lsm}
\end{center}
\end{figure}

The LSM is equivalent to the 
square-root measurement 
\cite{BKMO97,HW94,H96,SKIH98,SUIH98,KOSH99}, and 
has many desirable
properties. Its construction is relatively 
simple; it can be determined directly
from the given collection of states; it minimizes the probability
of a detection error for pure-state ensembles that exhibit certain symmetries
\cite{BKMO97,EF01}; it is ``pretty good''  when the states to be
distinguished are equally likely and almost orthogonal \cite{HW94};
it achieves a probability of error within a factor of two of the optimal probability of error \cite{BK00};
and it is asymptotically optimal \cite{H96,H98}.
Because of these properties, the LSM has been proposed as a detection
measurement in many applications (see \eg
\cite{SKIH98,SUIH98,KOSH99}).

It turns out that in many cases of practical interest the LSM is
optimal, \ie it minimizes the probability of a detection error.  From
the necessary and sufficient conditions for optimality discussed in
Section~\ref{sec:qd} it follows that the LSM minimizes the probability
of a detection error if and only if the measurement operators
$\hpi_i=\mu_i\mu_i^*$ defined by (\ref{eq:glsm}) satisfy
(\ref{eq:condz}) for some Hermitian $\hx$ satisfying (\ref{eq:condx}).
A sufficient condition for optimality of the LSM is given in the
following theorem, the proof of which is provided in the Appendix.
\begin{theorem}
\label{thm:gcondition}
Let $\{\rho_i=\phi_i\phi_i^*,1 \leq i \leq m\}$ denote a collection of quantum
states with prior probabilities $\{p_i,1 \leq i \leq m\}$. Let
$\{\Sigma_i=\mu_i\mu_i^*,1 \leq i \leq m\}$ with $\{\mu_i=T\psi_i,1
\leq i \leq m\}$
denote the
least-squares measurement (LSM) operators  corresponding to
$\{\psi_i=\sqrt{p_i}\phi_i,1 \leq i \leq m\}$,
where $T=(\Psi\Psi^*)^{-1/2}$ and $\Psi$ is the matrix with block columns
$\psi_i$. Then the LSM minimizes the probability of a detection
error if for each $i$,
$\mu_i^*\psi_i=\psi_i^*T\psi_i=\alpha I$, where
$\alpha$ is a constant independent of $i$. 
\end{theorem}
\noindent A similar result for the special case in which the density
operators $\rho_i$ are rank-one operators of the form
$\rho_i=\ket{\phi_i}\bra{\phi_i}$ and the vectors
$\ket{\phi_i}$ are linearly independent was derived in \cite{SKIH98}.

Note that the condition $\psi_i^*T\psi_i=\alpha I$ does not depend on
the choice of factor $\phi_i$. Indeed, if $\phi_i'=\phi_i Q_i$ is
another factor of $\rho_i$ with $Q_i$ satisfying $Q_iQ_i^*=I$, 
and if $\Psi'$ is the matrix of block columns
$\psi_i'=\sqrt{p_i}\phi'_i=\sqrt{p_i}\phi_iQ_i$, then 
it is easy to see that
$(\psi'_i)^*(\Psi' \Psi'^*)^{-1/2}\psi_i'=\alpha I$ if and only if 
$\psi_i^*T\psi_i=\alpha I$.

If the state $\rho_i=\phi_i\phi_i^*$ is transmitted
with prior probability $p_i$, then the probability of correctly
detecting the state
using measurement
operators $\Sigma_i=\mu_i\mu_i^*$ is
$p_i \tr(\mu_i^*\phi_i\phi_i^*\mu_i)=
\tr(\mu_i^*\psi_i\psi_i^*\mu_i)$.
It follows that if the condition for optimality of
Theorem~\ref{thm:gcondition} is met, so that $\mu_i^*\psi_i=\alpha I$, 
then the probability of correctly detecting each of the states
$\rho_i$ using the LSM is $\alpha^2$, independent of $i$. 

For a pure-state ensemble consisting of states $\ket{\phi_i}$ with prior
probabilities $p_i$, the probability of correct detection of the $i$th
state is given by $|\braket{\mu_i}{\psi_i}|^2$.  Since
$\braket{\mu_i}{\psi_i}=\braket{\psi_i}{T|\psi_i}>0$ for any set of
weighted vectors $\ket{\psi_i}$, $\braket{\mu_i}{\psi_i}$ is 
constant for all $i$ if and only if $|\braket{\mu_i}{\psi_i}|^2$ is
constant for all $i$.  Therefore, we may
interpret the condition in Theorem~\ref{thm:gcondition} for pure-state
ensembles as follows: The LSM is optimal for a set of states
$\ket{\phi_i}$ with prior probabilities $p_i$ if the probability of
detecting each one of the states using the LSM vectors is the same,
regardless of the specific state chosen.

In the remainder of the paper we use Theorem~\ref{thm:gcondition} to
derive the optimal measurement for mixed and pure state sets with
certain symmetry 
properties. The symmetry properties we consider are quite general,
and include many cases of practical interest.

%%%%%%%%%%%%%%%%%%%%%%%%%%%%%
\section{Geometrically Uniform State Sets}
\label{sec:gu}
%%%%%%%%%%%%%%%%%%%%%%%%%%

In this section we  consider the case in which the 
density operators $\rho_i$ are defined over a (not
necessarily abelian) group of
unitary matrices and are generated by a single generating matrix. Such a
state set is called  {\em geometrically uniform (GU)} \cite{F91}.
We first obtain a convenient characterization of the LSM in this case
and then show that under a certain constraint on the generator, the LSM
minimizes the probability of a detection error. In particular, for
pure-state ensembles the LSM minimizes the probability of a detection error.

Let $\G=\{U_i,1 \leq i \leq m\}$ be a finite  group of $m$ unitary matrices
$U_i$. That is, $\G$ contains the identity matrix $I$;
if $\G$ contains $U_i$, then it also contains its inverse
$U_i^{-1} = U_i^*$;  and the product $U_i U_j$ of any two elements of
$\G$ is in $\G$ \cite{A88}. 

A state set generated by $\G$ using a single generating operator
$\rho$ is a set $\SSS = \{\rho_i=U_i\rho U_i^*, U_i \in \G\}$.
The group $\G$ will be called the \emph{generating group} of
$\SSS$. For concreteness we assume that
$U_1=I$ so that $\rho_1=\rho$.
Such a state set has strong symmetry properties and 
is called GU.
For consistency with the symmetry of
$\SSS$, we will assume equiprobable prior probabilities on $\SSS$.

If the state set $\{\rho_i,1 \leq i \leq m\}$ is GU, then we can
always choose factors  $\phi_i$ of $\rho_i$ such that 
$\{\phi_i=U_i \phi,U_i \in \G\}$ where $\phi$ is a factor of $\rho$,
so that the 
factors $\phi_i$ are also GU with generator $\phi$.
In the remainder of this section we explicitly assume that the factors
are chosen to be GU.

We note that in \cite{EF01} a GU state set was defined for the case of
rank-one ensembles. Furthermore, the generating group was assumed to be
{\em abelian}.

In the next section we derive the LSM operators for GU state sets and 
show that the LSM operators are also GU with the
same generating group. We will see that this implies that when
using the LSM, the 
probability of 
correct detection of each of the states in a GU state set is the same
regardless of 
the particular state chosen. From
Theorem~\ref{thm:gcondition} it then follows that for pure-state
ensembles,  the LSM is optimal.

%%%%%%%%%%%%%%%%%%%%%%%%%%%%%
\subsection{The LSM for GU States}
%%%%%%%%%%%%%%%%%%%%%%%%%%

To derive the LSM for a GU state set with generating group $\G$, we
first show that $\Phi\Phi^*$ commutes
with each of the matrices $U_i \in \G$.  
Indeed,
expressing $\Phi\Phi^*$ as
\begin{equation}
\Phi\Phi^*=\sum_{i=1}^m\phi_i\phi_i^*=
\sum_{i=1}^m U_i\phi\phi^*U_i^*,
\end{equation}
we have that for all $j$,
\begin{eqnarray}
\Phi\Phi^* U_j & = & \sum_{i=1}^m
U_i\phi\phi^* U_i^*U_j \nonumber \\ 
& = & U_j \sum_{i=1}^m U_j ^*U_i\phi\phi^* U_i^*U_j \nonumber \\
& = & U_j \sum_{i=1}^m U_i\phi\phi^* U_i \nonumber \\
& = & U_j\Phi\Phi^*,
\end{eqnarray}
since $\{U_j ^*U_i,1 \leq i \leq m\}$ is just a permutation of $\G$.

If $\Phi\Phi^*$ commutes with $U_j$, then
\begin{equation}
\label{eq:M}
M=(\Phi\Phi^*)^{-1/2}
\end{equation}
also commutes
with $U_j$ for all $j$. Thus, from (\ref{eq:lsmep}) the LSM operators
are $\Sigma_i=\mu_i\mu_i^*$ with
\begin{equation}
\mu_i=M\phi_i=M U_i\phi=U_iM\phi=U_i \mu,
\end{equation}
where 
\begin{equation}
\label{eq:mu}
\mu=M \phi=(\Phi\Phi^*)^{-1/2} \phi.
\end{equation}
It follows that the LSM factors $\mu_i$ are also GU with generating
group $\G$ and 
generator $\mu$ given by (\ref{eq:mu}).
Therefore, to compute the LSM factors for a GU state set all we
need is to compute the generator $\mu$.
The remaining measurement factors are then obtained by applying the
group $\G$ to $\mu$.

A similar result was developed in \cite{EF01} for rank-one ensembles
in the  case in
which the group $\G$ is abelian using the Fourier transform defined
over $\G$. 

%%%%%%%%%%%%%%%%%%%%%%%%%%%%%
\subsection{Optimality of the LSM}
%%%%%%%%%%%%%%%%%%%%%%%%%%

We have seen that for a GU state set $\{\rho_i=\phi_i\phi_i^*,1 \leq i
\leq m\}$ with equal prior probabilities $1/m$ and generating group
$\G=\{U_i,1 \leq i \leq m\}$, the LSM operators
$\{\Sigma_i=\mu_i\mu_i^*,1 \leq i \leq m\}$ are also GU with
generating group $\G$. Therefore,
\begin{equation}
\label{eq:ip}
\mu_i^*\psi_i=\frac{1}{\sqrt{m}}\mu_i^*\phi_i=
\frac{1}{\sqrt{m}}\mu^*U_i^*U_i\phi=
\frac{1}{\sqrt{m}}\mu^*\phi,
\end{equation}
where $\phi$ and $\mu$ are the generators of the
the state factors and the LSM factors, respectively.
It follows that the probability of correct detection of each one of
the states $\rho_i$ using the LSM is the same, regardless of the state
transmitted. This then implies from Theorem~\ref{thm:gcondition} that
for pure-state GU ensembles the LSM is optimal. For a mixed-state ensemble,
if the generator $\phi$ satisfies 
\begin{equation}
\label{eq:gugc}
\mu^*\phi=\phi^*(\Phi\Phi^*)^{-1/2}\phi=\alpha I,
\end{equation}
for some $\alpha$, then
from Theorem~\ref{thm:gcondition} the 
LSM minimizes the probability of a detection error.

Note that the condition $\mu^*\phi=\alpha I$ does not depend on
the choice of generator $\phi$. Indeed, if $\phi'=\phi Q$ is
another factor of $\rho$, then from (\ref{eq:mu}) the generator of the
LSM factors is $\mu'=\mu Q$ 
so that  $\mu'^*\phi'=\alpha I$ if and only if 
$\mu^*\phi=\alpha I$.

%%%%%%%%%%%%%%%%%%%%%%%%%%%%%
\subsection{Optimal Measurement for Arbitrary GU State Sets}
\label{sec:agu}
%%%%%%%%%%%%%%%%%%%%%%%%%%

If the generator $\phi$ does not satisfy (\ref{eq:gugc}), then
the LSM is no longer guaranteed to be optimal. Nonetheless, as we now
show, the optimal measurement operators that minimize the probability
of a detection error are GU with generating group $\G$. The
corresponding generator can be computed very efficiently in polynomial
time.

Suppose that the optimal measurement operators  that maximize 
\begin{equation}
J(\{\Pi_i\})=\sum_{i=1}^m \tr(\rho_i \Pi_i),
\end{equation}
are $\hpi_i$, and let
$\widehat{J}=J(\{\hpi_i\})=\sum_{i=1}^m\tr(\rho_i \hpi_i)$.
Let $r(j,i)$ be the mapping from $\I \times \I$ to $\I$ with
$\I=\{1,\ldots,m\}$, defined by
$r(j,i)=k$ if 
$U_j^*U_i=U_k$.
Then the measurement operators 
$\hpi_i'=U_j\hpi_{r(j,i)}U_j^*$ for any $1 \leq j \leq m$ are also
optimal.   
Indeed, since $\hpi_i \geq 0$ and $\sum_{i=1}^m \hpi_i=I$, $\hpi'_i
\geq 
0$ and
\begin{equation}
\sum_{i=1}^m \hpi'_i=U_j\bl \sum_{i=1}^m \hpi_i \br U_j^*=U_jU_j^*=I.
\end{equation}
Finally, using the fact that $\rho_i=U_i\rho U_i^*$ for some generator
$\rho$, 
\begin{eqnarray}
J(\{\hpi'_i\})=
\sum_{i=1}^m\tr(\rho U_i^* U_j\hpi_{r(j,i)}U_j^*U_i)=
\sum_{k=1}^m\tr(\rho U_k^*\hpi_kU_k)=
\sum_{i=1}^m\tr(\rho_i\hpi_i)=\widehat{J}.
\end{eqnarray}

Since the measurement operators $\{\hpi_i'=U_j\hpi_{r(j,i)}U_j^*,1 
\leq i \leq m\}$ are 
optimal for any $j$, it 
follows immediately that the measurement operators
$\{\overline{\Pi}_i=(1/m)\sum_{j=1}^m 
U_j\hpi_{r(j,i)}U_j^*,1 \leq i \leq m\}$ are also optimal.
Indeed, it is immediate that $\overline{\Pi}_i$ 
satisfy 
(\ref{eq:identu}). In 
addition,
$J(\{\overline{\Pi}_i\})=J(\{\hpi'_i\})=\widehat{J}$. 
Now,
\begin{eqnarray}
\overline{\Pi}_i & = & \frac{1}{m}\sum_{j=1}^m U_j\hpi_{r(j,i)}U_j^*
\nonumber \\
 & = & \frac{1}{m}\sum_{k=1}^m U_iU_k^*\hpi_kU_kU_i^* \nonumber \\
 & = & U_i \bl \frac{1}{m}\sum_{k=1}^mU_k^*\hpi_kU_k \br U_i^* 
\nonumber \\ 
 & = & U_i \widehat{\Pi} U_i^*, 
\end{eqnarray}
where $\widehat{\Pi}=(1/m)\sum_{k=1}^mU_k^*\hpi_kU_k$. 

We therefore conclude that the optimal measurement operators can always
be chosen to be GU with the same generating group $\G$ as the original
state set. Thus, to find the optimal measurement operators all we need
is to find 
the optimal generator  $\hpi$. The remaining  operators are
obtained by applying the group $\G$ to $\hpi$.

Since the optimal measurement operators satisfy $\Pi_i=U_i \Pi U_i^*$
and $\rho_i=U_i \rho U_i^*$, $\tr(\rho_i \Pi_i)=\tr (\rho \Pi)$, so that the
problem (\ref{eq:pe}) reduces to
the maximization problem
\begin{equation}
\label{eq:max}
\max_{\Pi \in \B} \tr(\rho\Pi),
\end{equation} 
where $\B$ is
the set of Hermitian operators on $\HH$,
subject to the constraints
\begin{eqnarray}
\label{eq:condp}
\Pi & \geq & 0; \nonumber \\
\sum_{i=1}^m U_i \Pi U_i^* & = & I.
\end{eqnarray}
Since this problem is a (convex) semidefinite programming problem, the
optimal $\Pi$ 
can be computed very efficiently in polynomial
time within any desired accuracy \cite{VB96,A91t,NN94}, for example
using the LMI toolbox on Matlab.
Note that the problem of (\ref{eq:max}) and (\ref{eq:condp}) has $n^2$
real unknowns and 
$2$ constraints, in
contrast 
with the original maximization problem (\ref{eq:pe}) and
(\ref{eq:identu}) which has $mn^2$
real unknowns 
and $m+1$ constraints.

We summarize our results regarding GU state sets in the following theorem:
\begin{theorem}[GU state sets]
\label{thm:gu}
Let $\SSS = \{\rho_i =
U_i\rho U_i^*, U_i \in \G\}$
be a geometrically uniform (GU) state set generated
by a finite group $\G$ of unitary matrices, where
$\rho=\phi\phi^*$ is an arbitrary generator, and let $\Phi$ be the matrix of
columns $\phi_i=U_i\phi$.  
Then the
least-squares measurement (LSM) is given by the
measurement operators $\Sigma_i=\mu_i\mu_i^*$ with
\[\mu_i=U_i\mu\]
where
\[\mu=(\Phi\Phi^*)^{-1/2} \phi.\]
The LSM has the following properties:
\begin{enumerate}
\item The measurement operators are GU with generating group $\G$;
\item The probability of correctly detecting each of the states
$\rho_i$ using the LSM is the same;
\item If $\mu^*\phi=\phi^*(\Phi\Phi^*)^{-1/2} \phi=\alpha I$
then the LSM minimizes the probability of a
detection error; In particular, if $\phi=\ket{\phi}$ is a vector so
that the 
state set is a rank-one 
ensemble, then the LSM minimizes the probability of a detection error.
\end{enumerate}
For an arbitrary generator $\phi$ the optimal measurement operators
that minimize the probability of a detection error are also 
GU with generating group $\G$ and generator $\Pi$ that maximizes
$\tr(\rho \Pi)$ subject to $\Pi \geq 0$ and 
$\sum_{i=1}^m U_i \Pi U_i^* = I$.
\end{theorem}

%%%%%%%%%%%%%%%%%%%%%%%%%%%%%
\section{Compound Geometrically Uniform State Sets}
\label{sec:cgu}
%%%%%%%%%%%%%%%%%%%%%%%%%%

In this section, we consider state sets which consist of
subsets that are GU, and are therefore referred to as {\em compound
geometrically uniform (CGU)} \cite{EB01}. As we show, the LSM operators
are also CGU so that they can be computed using a {\em set} of
generators.  Under a certain condition on the generators, we
also show that the optimal measurement associated with a CGU state set is equal to
the LSM. For arbitrary CGU state sets, the optimal measurement is no longer equal to
the LSM. Nonetheless, as we show, the optimal measurement operators
are also CGU and we derive an efficient computational method for 
finding the optimal generators.

A CGU state set is defined as a set of density operators 
$\SSS=\{\rho_{ik}=\phi_{ik}\phi_{ik}^*,1 \leq i \leq l,1 \leq k \leq
r\}$ such that
$\rho_{ik}=U_i\rho_kU_i^*$, where the matrices $\{U_i,1 \leq i
\leq l\}$ are unitary and form a
 group $\G$, and the operators $\{\rho_k,1 \leq k \leq r\}$  are
the generators.
For concreteness we assume that
$U_1=I$ so that $\rho_{1k}=\rho_k$.
We also assume equiprobable prior probabilities on $\SSS$.

If the state set $\{\rho_{ik},1 \leq i \leq l,1 \leq k \leq r\}$ is
CGU, then we can 
always choose factors  $\phi_{ik}$ of $\rho_{ik}$ such that 
$\{\phi_{ik}=U_i \phi_k,1 \leq i \leq l\}$ where $\phi_k$ is a factor
of $\rho_k$, 
so that the 
factors $\phi_{ik}$ are also CGU with generators $\{\phi_k,1 \leq k
\leq r\}$.
In the remainder of this section we explicitly assume that the factors
are chosen to be CGU.
  
A CGU state set  is in general not GU. However,
for every $k$, the matrices $\{\phi_{ik},1 \leq i \leq l\}$ and the
operators $\{\rho_{ik},1 \leq i \leq l\}$ are GU
with generating group $\G$.

\newpage
%%%%%%%%%%%%%%%%%%%%%%%%%%%%%
\subsection{Example of a Compound Geometrically Uniform State Set}
\label{sec:cgue}
%%%%%%%%%%%%%%%%%%%%%%%%%%

An example of a CGU state set is illustrated in Fig.~\ref{fig:cgu}.
In this example the state set is
$\{\rho_{ik}=\ket{\phi_{ik}}\bra{\phi_{ik}},1 \leq i,k
\leq 2\}$ where  
$\{\ket{\phi_{ik}}=U_i\ket{\phi_k},U_i \in \G\}$, $\G=\{I_2,U\}$ with
\begin{equation}
\label{eq:U}
U=\frac{1}{2}\left[
\begin{array}{rr}
1 & \sqrt{3}  \\
\sqrt{3} & -1 
\end{array}
\right],
\end{equation}
and the generating vectors are 
\begin{equation}
\ket{\phi_1}=\frac{1}{\sqrt{2}}\left[
\begin{array}{r}
1  \\
1
\end{array}
\right],\,\,\,
\ket{\phi_{2}}=\frac{1}{\sqrt{2}}\left[
\begin{array}{r}
1  \\
-1
\end{array}
\right].
\end{equation}
The matrix $U$ represents a reflection about the dashed line in
Fig.~\ref{fig:cgu}. Thus, the vector $\ket{\phi_{21}}$ is obtained by
reflecting the generator $\ket{\phi_{11}}$ about this line, and
similarly the vector $\ket{\phi_{22}}$ is obtained by
reflecting the generator $\ket{\phi_{12}}$ about this line.

As can be seen from the figure, the state
set is not GU. In particular, there is no isometry
that transforms $\ket{\phi_{11}}$ into $\ket{\phi_{12}}$ while leaving the set
invariant. However, the sets 
$\SSS_1=\{\ket{\phi_{11}},\ket{\phi_{21}}\}$
and $\SSS_2=\{\ket{\phi_{12}},\ket{\phi_{22}}\}$ are both GU with
generating group $\G$.
\setlength{\unitlength}{.2in}
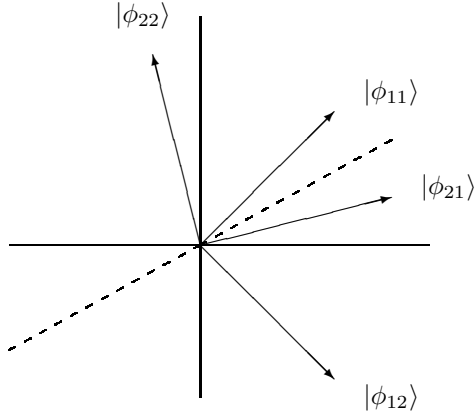
\begin{figure}[h]
\begin{center}
\begin{picture}(15,11)(0,0)
%\put(0,0){\framebox(15,11){}}

\put(0,-1.5){
\put(7,6){\line(0,1){6}}
\put(7,6){\line(1,0){6}}
\put(7,6){\line(0,-1){4}}
\put(7,6){\line(-1,0){5}}

\put(7,6){\vector(1,1){3.5}}
\put(7,6){\vector(1,-1){3.5}}
\put(7,6){\vector(4,1){5}}
\put(7,6){\vector(-1,4){1.25}}

\put(12,10){\makebox(0,0){\small $\ket{\phi_{11}}$}}
\put(12,2){\makebox(0,0){\small $\ket{\phi_{12}}$}}
\put(13.5,7.5){\makebox(0,0){\small $\ket{\phi_{21}}$}}
\put(5.5,12){\makebox(0,0){\small $\ket{\phi_{22}}$}}

\dashline{.2}(7,6)(12,8.75)
\dashline{.2}(7,6)(2,3.25)

}
\end{picture}
\caption{A compound geometrically uniform pure-state set. The state sets
$\SSS_1=\{\ket{\phi_{11}},\ket{\phi_{21}}\}$
and $\SSS_2=\{\ket{\phi_{12}},\ket{\phi_{22}}\}$ are both geometrically uniform
(GU) with the same generating group; Both sets 
are invariant under a reflection about the dashed line.
However, the combined set
$\SSS=\{\ket{\phi_{11}},\ket{\phi_{21}},\ket{\phi_{12}},\ket{\phi_{22}}\}$
is no longer GU. }
\label{fig:cgu}
\end{center}
\end{figure}

%%%%%%%%%%%%%%%%%%%%%%%%%%%%%
\subsection{The LSM for CGU State Sets}
%%%%%%%%%%%%%%%%%%%%%%%%%%

We now derive the LSM for a CGU state set with equal prior probabilities.
Let $\Phi$ denote the matrix of columns $\phi_{ik}$.
Then 
for a CGU state set with generating group $\G$,  $\Phi\Phi^*$ commutes
with each of the matrices $U_i \in \G$.  
Indeed,
expressing $\Phi\Phi^*$ as
\begin{equation}
\Phi\Phi^*=\sum_{i=1}^l\sum_{k=1}^r\phi_{ik}\phi_{ik}=
\sum_{i=1}^l U_i\bl \sum_{k=1}^r \phi_k\phi_k\br U_i^*,
\end{equation}
we have that for all $j$,
\begin{eqnarray}
\Phi\Phi^* U_j & = &
\sum_{i=1}^l U_i\bl \sum_{k=1}^r \phi_k\phi_k\br U_i^*U_j
\nonumber \\ 
& = & U_j 
\sum_{i=1}^l U_j^*U_i\bl \sum_{k=1}^r \phi_k\phi_k\br U_i^*U_j
\nonumber \\
& = & U_j 
\sum_{i=1}^l U_i\bl \sum_{k=1}^r \phi_k\phi_k\br U_i^*
\nonumber \\
& = & U_j\Phi\Phi^*,
\end{eqnarray}
since $\{U_j ^*U_i,1 \leq i \leq l\}$ is just a permutation of $\G$.

If $\Phi\Phi^*$ commutes with $U_j$, then
$M=(\Phi\Phi^*)^{-1/2}$ also commutes
with $U_j$ for all $j$. Thus, the  LSM operators are
$\Sigma_{ik}=\mu_{ik}\mu_{ik}^*$ with
\begin{equation}
\mu_{ik}=M\phi_{ik}=M U_i\phi_k=U_iM\phi_k=U_i \mu_k,
\end{equation}
where 
\begin{equation}
\label{eq:muc}
\mu_k=M \phi_k=(\Phi\Phi^*)^{-1/2} \phi_k.
\end{equation}
Therefore the LSM factors are also CGU with generating group $\G$ and
generators $\mu_k$ given by (\ref{eq:muc}).
To compute the LSM factors all we
need is to compute  the generators $\mu_k$.
The remaining measurement factors are then obtained by applying the
group $\G$ to each of the generators.

For the CGU state set of Fig.~\ref{fig:cgu} we have that
\begin{equation}
\Phi\Phi^*=2\left[
\begin{array}{rr}
1 & 0  \\
0 & 1 
\end{array}
\right].
\end{equation}
Therefore, the LSM  vectors are $\{\ket{\mu_{ik}}=U_i\ket{\mu_k},U_i \in
\G,1 \leq i,k \leq 2\}$ where $\G=\{I_2,U\}$ with $U$ given by
(\ref{eq:U}), and from (\ref{eq:muc}) the generating vectors are 
\begin{equation}
\ket{\mu_1}=
\frac{1}{\sqrt{2}}\ket{\phi_1}=
\frac{1}{2}\left[
\begin{array}{r}
1  \\
1
\end{array}
\right],\,\,\,
\ket{\mu_{2}}=
\frac{1}{\sqrt{2}}\ket{\phi_2}=
\frac{1}{2}\left[
\begin{array}{r}
1  \\
-1
\end{array}
\right].
\end{equation}
The LSM vectors are depicted in Fig.~\ref{fig:cgulsm}. The vectors
have the same symmetries as the state set of Fig.~\ref{fig:cgu}, with
different generating vectors. Since in this example the generating
vectors satisfy $\ket{\mu_k}=(1/\sqrt{2})\ket{\phi_k}$, we have that
$\ket{\mu_{ik}}=(1/\sqrt{2})\ket{\phi_{ik}}$ for $1 \leq i,k \leq 2$.
\setlength{\unitlength}{.2in}
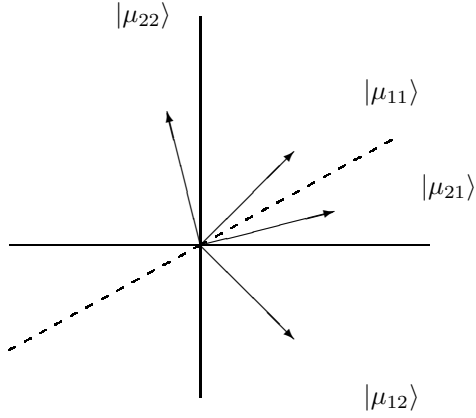
\begin{figure}[h]
\begin{center}
\begin{picture}(15,11)(0,0)
%\put(0,0){\framebox(15,11){}}

\put(0,-1.5){
\put(7,6){\line(0,1){6}}
\put(7,6){\line(1,0){6}}
\put(7,6){\line(0,-1){4}}
\put(7,6){\line(-1,0){5}}

\put(7,6){\vector(1,1){2.45}}
\put(7,6){\vector(1,-1){2.45}}
\put(7,6){\vector(4,1){3.5}}
\put(7,6){\vector(-1,4){0.875}}

\put(12,10){\makebox(0,0){\small $\ket{\mu_{11}}$}}
\put(12,2){\makebox(0,0){\small $\ket{\mu_{12}}$}}
\put(13.5,7.5){\makebox(0,0){\small $\ket{\mu_{21}}$}}
\put(5.5,12){\makebox(0,0){\small $\ket{\mu_{22}}$}}

\dashline{.2}(7,6)(12,8.75)
\dashline{.2}(7,6)(2,3.25)

}
\end{picture}
\caption{The least-squares measurement vectors associated with
the compound geometrically uniform state set of Fig.~\ref{fig:cgu}. 
As can be seen from the figure, the measurement vectors have the same
symmetries as the original state 
set. }
\label{fig:cgulsm}
\end{center}
\end{figure}

%%%%%%%%%%%%%%%%%%%%%%%%%%%%%%%%%%%%%%%%
\subsection{CGU State Sets  With GU Generators}
\label{sec:cgugu}
%%%%%%%%%%%%%%%%%%%%%%%%%%%%%%%%%%%%%%

A special class of CGU state sets  is {\em CGU
state sets  with GU generators} in which the generators
$\{\rho_k=\phi_k\phi_k^*,1 
\leq k \leq r\}$ and the factors $\phi_k$ are 
themselves GU. Specifically, $\{\phi_k=V_k \phi\}$ for
some generator $\phi$, 
where the matrices $\{V_k,1 \leq k \leq r\}$ are unitary, and form a
group $\Q$.

Suppose that $U_p$ and $V_t$ commute up to a phase factor for all
$t$ and $p$ so that $U_pV_t=V_tU_pe^{j\theta(p,t)}$ where $\theta(p,t)$
is an arbitrary phase function that may depend on the indices $p$ and
$t$. In this case we say that $\G$ and $\Q$ commute up to a phase
factor and that the corresponding state set is {\em CGU with commuting GU
generators}.  
(In the  special case in which $\theta=0$ so that
$U_iV_k=V_kU_i$ for 
all $i,k$, the resulting state set is GU \cite{EB01}).
Then for  all $p,t$,
\begin{eqnarray}
\Phi\Phi^*U_pV_t & = & \sum_{i=1}^lU_i \bl \sum_{k=1}^rV_k\phi \phi^*V_k^* \br U_i^*U_pV_t
\nonumber \\
& = & U_pV_t\sum_{i=1}^lV_t^*U_p^*U_i \bl \sum_{k=1}^rV_k \phi \phi^*V_k^*\br
U_i^*U_pV_t 
\nonumber \\
& = & U_p V_t \sum_{i=1}^lV_t^*U_i \bl \sum_{k=1}^rV_k \phi \phi^*V_k^*\br
U_i^* V_t 
\nonumber \\
& = & U_p V_t \sum_{i=1}^lU_i \bl \sum_{k=1}^rV_t^* V_k \phi \phi^*V_k^*V_t\br U_i^* 
\nonumber \\
& = & U_p V_t \sum_{i=1}^lU_i \bl \sum_{k=1}^rV_k\phi \phi^*V_k^*\br U_i^* 
\nonumber \\
& = & U_pV_t \Phi\Phi^*.
\end{eqnarray}
The LSM factors $\mu_{ik}$ are then given by
\begin{equation}
\label{eq:lsmcgugu}
\mu_{ik}= M\phi_{ik}=M U_iV_k \phi=U_iV_kM
\phi=U_i V_k\bar{\mu}, 
\end{equation}
where $\bar{\mu}=M \phi$.
Thus even though the state set is not in
general GU, the LSM factors can be computed using
a single generator.

Alternatively, we can express $\mu_{ik}$ as
$\mu_{ik}=U_i\mu_k$ where the generators $\mu_k$ are
given by
\begin{equation}
\label{eq:lsmg}
\mu_k=V_k\bar{\mu}.
\end{equation} 
From (\ref{eq:lsmg}) it follows that the generators $\mu_k$ are GU with
generating group $\Q=\{V_k,1 \leq k \leq r\}$ and generator
$\bar{\mu}$. 

We conclude that for a CGU state set with commuting GU generators and
generating group $\Q$, the LSM
vectors are also CGU  with commuting GU generators and generating
group $\Q$.

%%%%%%%%%%%%%%%%%%%%%%%%%%%%%%%%%%%%%%%%%
\subsection{Example of a CGU State Set with Commuting GU Generators}
%%%%%%%%%%%%%%%%%%%%%%%%%%%%%%%%%%%%%%%

We now consider an example of a CGU state set with commuting
GU generators. 
Consider the group $\G$ of $l$ unitary matrices on $\C^l$ where
$U_i=Z^i,1 \leq i \leq l$ and $Z$ is the matrix defined by 
\begin{equation}
Z\left[
\begin{array}{c}
x_1  \\
x_2  \\
\vdots \\
x_{l-1} \\
x_l
\end{array}
\right]=
\left[
\begin{array}{c}
x_2  \\
x_3  \\
\vdots \\
x_l\\
x_1
\end{array}
\right].
\end{equation}
Let $\Q$ be the group of $r=l$ unitary matrices $V_k=B^k,1 \leq k
\leq l$ where $B$ is the diagonal matrix with diagonal elements
$e^{j2\pi s/l},0 \leq s \leq l-1$. We can immediately verify that for
this choice of $\Q$ and $\G$, $U_iV_k=V_kU_ie^{j2\pi/l}$. We therefore
conclude that the LSM operators corresponding to the CGU state set
$\SSS=\{\rho_{ik}=\phi_{ik}\phi_{ik}^*\}$ with $\{\phi_{ik}=
U_i\phi_k,U_i \in \G\}$ and
$\{\phi_k=V_k \phi,V_k \in \Q\}$ 
 for some generator
$\phi$, are also CGU with commuting GU generators and can
therefore be
generated by a single generator.

As a special case, suppose that $l=2$ so that $\G$ consists of the
matrices 
$U_1=I_2$ and $U_2=Z$ where 
\begin{equation}
\label{eq:Z}
Z=\left[
\begin{array}{rr}
0 & 1 \\
1 & 0  \\
\end{array}
\right],
\end{equation}
and $\Q$ consists of the matrices
$V_1=I_2$ and $V_2=B$ where
\begin{equation}
\label{eq:B}
B=\left[
\begin{array}{rr}
1 & 0 \\
0 & -1  \\
\end{array}
\right].
\end{equation}
Let the state set be $\SSS=\{\ket{\phi_{ik}}=U_iV_k\ket{\phi},\,\,1
\leq i,k \leq 
2\}$,  where $\phi=[\beta_1\,\,\, \beta_2]^*$. Since
$\ket{\phi}$ must be normalized, $\beta_1^2+\beta_2^2=1$.
Then,
\begin{equation}
\label{eq:cguex}
\ket{\phi_{11}}=\left[
\begin{array}{r}
\beta_1  \\
\beta_2  
\end{array}
\right],\,\,\,
\ket{\phi_{21}}=\left[
\begin{array}{r}
\beta_2  \\
\beta_1  
\end{array}
\right],\,\,\,
\ket{\phi_{12}}=\left[
\begin{array}{r}
\beta_1  \\
-\beta_2  
\end{array}
\right],\,\,\,
\ket{\phi_{22}}=\left[
\begin{array}{r}
-\beta_2  \\
\beta_1  
\end{array}
\right].
\end{equation}
The LSM vectors are given by 
$\{\ket{\mu_{ik}}=U_iV_k\ket{\bar{\mu}},\,\,1 \leq i,k \leq
2\}$,  where $\ket{\bar{\mu}}=(\Phi\Phi^*)^{-1/2}\ket{\phi}$, and
\begin{equation}
\label{eq:phi}
\Phi =\left[
\begin{array}{rrrr}
\beta_1 &  \beta_2 &  \beta_1 &  -\beta_2  \\
\beta_2 &  \beta_1 &  -\beta_2 &  \beta_1 
\end{array}
\right].
\end{equation}
We can immediately verify that 
\begin{equation}
\Phi\Phi^* =\left[
\begin{array}{cc}
2(\beta_1^2+\beta_2^2) &  0  \\
0 &  2(\beta_1^2+\beta_2^2)
\end{array}
\right]=
\left[
\begin{array}{cc}
2 &  0  \\
0 &  2
\end{array}
\right].
\end{equation}
Thus, the LSM vectors are 
\begin{equation}
\ket{\mu_{ik}}=
\frac{1}{\sqrt{2}}\ket{\phi_{ik}}.
\end{equation}

In Fig.~\ref{fig:cguex} we plot the state vectors given by
(\ref{eq:cguex}) for the case in which
$\ket{\phi}=(1/\sqrt{5})[2\,\,\,1]^*$. As can be seen from the figure, the state
set is not GU. In particular, there is no isometry
that transforms $\ket{\phi_{11}}$ into $\ket{\phi_{12}}$ while leaving the set
invariant. 
Nonetheless, we have seen that the LSM vectors can be
generated by a single generating vector
$\ket{\mu}=(1/\sqrt{2})\ket{\phi}$.
\setlength{\unitlength}{.2in}
\begin{figure}[h]
\begin{center}
\begin{picture}(15,11)(0,0)
%\put(0,0){\framebox(15,11){}}

\put(0,-1.5){
\put(7,6){\line(0,1){6}}
\put(7,6){\line(1,0){6}}
\put(7,6){\line(0,-1){4}}
\put(7,6){\line(-1,0){5}}

\put(7,6){\vector(2,1){4}}
\put(7,6){\vector(1,2){2}}
\put(7,6){\vector(2,-1){4}}
\put(7,6){\vector(-1,2){2}}

\put(12.5,8){\makebox(0,0){\small $\ket{\phi_{11}}$}}
\put(12.5,4){\makebox(0,0){\small $\ket{\phi_{12}}$}}
\put(9,11){\makebox(0,0){\small $\ket{\phi_{21}}$}}
\put(5,11){\makebox(0,0){\small $\ket{\phi_{22}}$}}

\dashline{.2}(7,6)(12,11)
\dashline{.2}(7,6)(4,3)

}
\end{picture}
\caption{A compound geometrically uniform state set with commuting
geometrically uniform (GU) generators. The state sets
$\SSS_1=\{\ket{\phi_{11}},\ket{\phi_{21}}\}$ and
$\SSS_2=\{\ket{\phi_{12}},\ket{\phi_{22}}\}$ are both GU with the same
generating group; Both sets are invariant under a reflection about the
dashed line.  The set of generators
$\{\ket{\phi_{11}},\ket{\phi_{12}}\}$ is GU and is invariant under a
reflection about the $x$-axis.  The combined set
$\SSS=\{\ket{\phi_{11}},\ket{\phi_{21}},\ket{\phi_{12}},\ket{\phi_{22}}\}$
is no longer GU. Nonetheless, the LSM vectors are generated by a
single generating vector and are given by
$\ket{\mu_{ik}}=(1/\sqrt{2})\ket{\phi_{ik}}$.}
\label{fig:cguex}
\end{center}
\end{figure}
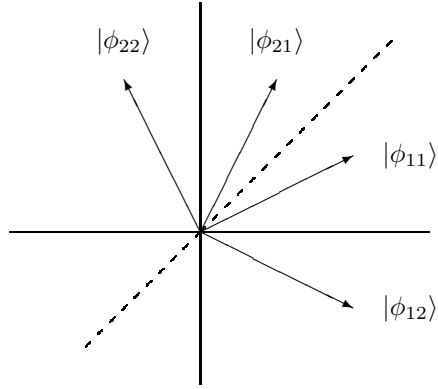

Note, that in the example of Section~\ref{sec:cgue} the CGU state set
also has
GU generators.
Specifically, the set of generators
$\{\ket{\phi_1},\ket{\phi_2}\}$ with
$\ket{\phi_1}=\ket{\phi_{11}}$ and $\ket{\phi_2}=\ket{\phi_{12}}$
is invariant under a reflection about the $x$-axis:
$\ket{\phi_2}=B\ket{\phi_1}$ where $B$ is given by (\ref{eq:B}).
However, the group $\Q=\{I_2,B\}$ of generators does not
commute up 
to a phase with the generating group $\G=\{I_2,U\}$, where 
$U$ is given by (\ref{eq:U}) and represents a reflection about
the dashed line in Fig.~\ref{fig:cgu}. 
This can be verified graphically from Fig.~\ref{fig:cgu}: Suppose we
apply $B$ to 
$\ket{\phi_{11}}$ and then apply $U$. Then the resulting vector is
equal to $\ket{\phi_{22}}$. If on the other hand we first apply $U$ to
$\ket{\phi_{11}}$ and then apply $B$, then the  resulting vector is the
reflection of  $\ket{\phi_{21}}$ about the $x$-axis, which is not
related to  $\ket{\phi_{22}}$ by a phase factor.

Now, consider the state set in Fig.~\ref{fig:cguex}. In this case 
$\Q=\{I_2,B\}$ and  $\G=\{I_2,Z\}$ where 
$B$ represents a
reflection about the $x$-axis and $Z$ represents a reflection about
the dashed line in Fig.~\ref{fig:cguex}. We can immediately verify  from
the figure that applying $Z$ and then $B$ to any vector in the set
results in a vector that is equal up to a minus sign to the vector
that results from first applying $B$ and then $Z$. For example,
applying $B$ to
$\ket{\phi_{11}}$ and then applying $Z$ results in
$\ket{\phi_{22}}$. If on the other hand we first apply $Z$ to
$\ket{\phi_{11}}$ and then apply $B$, then the resulting vector is the
reflection of  $\ket{\phi_{21}}$ about the $x$-axis, which is equal to
$-\ket{\phi_{22}}$.

%%%%%%%%%%%%%%%%%%%%%%%%%%%%%
\subsection{Optimality of the LSM}
%%%%%%%%%%%%%%%%%%%%%%%%%%

We have seen in the previous section that the LSM operators corresponding
to a CGU state set with generating group $\G=\{U_i,1 \leq i \leq l\}$
is also CGU with the same generating group.
In particular for each $k$, the sets
$\SSS'_k=\{\mu_{ik},1 \leq i \leq l\}$  and
$\SSS_k=\{\phi_{ik},1 \leq i \leq l\}$  are both GU with
generating group $\G$. Therefore, 
\begin{equation}
\label{eq:cgu_condt}
\mu_{ik}^*\phi_{ik}=\mu_k^*U_i^*U_i\phi_{k}=\mu_k^*\phi_{k},
\end{equation}
which implies that the probability of correctly detecting each of
the states in $\SSS_k$ using the LSM is the same.
It follows from Theorem~\ref{thm:gcondition}
that if 
\begin{equation}
\label{eq:cgu_cond}
\mu^*_k\phi_k=\phi_k^*M\phi_k=\alpha I,\quad 1 \leq k \leq r,
\end{equation}
then the LSM minimizes the probability of a detection error.

Note that the condition $\mu_k^*\phi_k=\alpha I$ does not depend on
the choice of generator $\phi_k$. Indeed, if $\phi_k'=\phi_k Q_k$ is
another factor of $\rho_k$, then from (\ref{eq:muc}) the generator of the
LSM factors is $\mu_k'=\mu_k Q_k$ 
so that  $\mu_k'^*\phi_k'=\alpha I$ if and only if 
$\mu_k^*\phi_k=\alpha I$.

In Section~\ref{sec:cgugu} we showed that the LSM operators
corresponding to a  CGU state set 
with GU generators 
$\{\phi_k=V_k \phi\}$ where $V_k \in \Q$ and 
$\G$ and $\Q$ commute up to a phase
factor,  are also CGU with
GU generators generated by the same group $\Q$ and some generator
$\bar{\mu}$.
Therefore for all $k$,
\begin{equation}
\label{eq:gc}
\mu_k^*\phi_k=
\bar{\mu}^*V_k^*V_k\phi=
\bar{\mu}^*\phi,
\end{equation}
so that the probability of correctly detecting each of the states
$\phi_{ik}$ is the same.
If in addition,
\begin{equation}
\label{eq:cmguc}
\bar{\mu}^*\phi=\phi^*M\phi=\alpha I, 
\end{equation}
then combining (\ref{eq:cgu_condt}), (\ref{eq:gc}) and
(\ref{eq:cmguc}) with Theorem~\ref{thm:gcondition} we conclude that
the LSM
minimizes the probability of a detection error.
In particular, for a rank-one ensemble, $\bar{\mu}^*\phi$ is a scalar
so that (\ref{eq:cmguc}) is always satisfied.
Therefore, for a rank-one CGU state set with commuting GU generators,
the LSM minimizes the probability of a detection error.

%%%%%%%%%%%%%%%%%%%%%%%%%%%%%
\subsection{Optimal Measurement for Arbitrary CGU State Sets}
%%%%%%%%%%%%%%%%%%%%%%%%%%

If the generators $\phi_k$ do not satisfy (\ref{eq:cgu_cond}), then
the LSM is no longer guaranteed to be optimal. Nonetheless, as we now
show, the optimal measurement operators that minimize the probability
of a detection error are CGU with generating group $\G$. The
corresponding generators can be computed very efficiently in polynomial
time within any desired accuracy.

Suppose that the optimal measurement operators  that maximize 
\begin{equation}
J(\{\Pi_{ik}\})=\sum_{i=1}^l \sum_{k=1}^r\tr(\rho_{ik} \Pi_{ik}),
\end{equation}
are $\hpi_{ik}$, and let $\widehat{J}=J(\{\hpi_{ik}\})=\sum_{i=1}^l
\sum_{k=1}^r \tr(\rho_{ik} \hpi_{ik})$.  Let $r(j,i)$ be the mapping
from $\I \times \I$ to 
$\I$ with $\I=\{1,\ldots,l\}$, defined by $r(j,i)=s$ if $U_j^*U_i=U_s$. 
Then the measurement
operators $\hpi_{ik}'=U_j\hpi_{r(j,i)k}U_j^*$ for any $1 \leq j \leq l$
are also optimal. 
Indeed, since $\hpi_{ik} \geq 0$ and $\sum_{i,k} \hpi_{ik}=I$,
$\hpi'_{ik} \geq 0$ and
\begin{equation}
\sum_{i=1}^l \sum_{k=1}^r \hpi'_{ik}=U_j\bl \sum_{i=1}^l \sum_{k=1}^r
\hpi_{ik} \br U_j^*=U_jU_j^*=I. 
\end{equation}
Finally, using the fact that $\rho_{ik}=U_i\rho_k U_i^*$ for some generators
$\rho_k$, 
\begin{eqnarray}
J(\{\hpi'_{ik}\})=
\sum_{i=1}^l \sum_{k=1}^r \tr(\rho_k U_i^* U_j\hpi_{r(j,i)k}U_j^*U_i)=
\sum_{s=1}^l \sum_{k=1}^r\tr(\rho_k U_s^*\hpi_{sk}U_s)=
\sum_{i=1}^l \sum_{k=1}^r\tr(\rho_{ik} \hpi_{ik})=\widehat{J}.
\end{eqnarray}

Since the measurement operators $\{\hpi'_{ik}=U_j\hpi_{r(j,i)k}U_j^*,1 
\leq i \leq l,1 \leq k \leq r\}$ are 
optimal for any $j$, it 
follows immediately that the measurement operators
$\{\overline{\Pi}_{ik}=(1/l)\sum_{j=1}^l 
U_j\hpi_{r(j,i)k}U_j^*,1 \leq i \leq l,1 \leq k \leq r\}$ are also
optimal. 
Indeed, it is immediate that $\overline{\Pi}_{ik}$ 
satisfy 
(\ref{eq:identu}). In 
addition,
$J(\{\overline{\Pi}_{ik}\})=J(\{\hpi'_{ik}\})=\widehat{J}$. 
Now,
\begin{eqnarray}
\overline{\Pi}_{ik} & = & \frac{1}{l}\sum_{j=1}^l U_j\hpi_{r(j,i)k}U_j^*
\nonumber \\
 & = & \frac{1}{l}\sum_{s=1}^l U_iU_s^*\hpi_{sk}U_sU_i^* \nonumber \\
 & = & U_i \bl \frac{1}{l}\sum_{s=1}^lU_s^*\hpi_{sk}U_s \br U_i^* 
\nonumber \\ 
 & = & U_i \widehat{\Pi}_k U_i^*, 
\end{eqnarray}
where $\widehat{\Pi}_k=(1/l)\sum_{s=1}^lU_s^*\hpi_{sk}U_s$. 

We therefore conclude that the optimal measurement operators can always
be chosen to be CGU with the same generating group $\G$ as the original
state set. Thus, to find the optimal measurement operators all we need
is to find 
the optimal generators  $\{\hpi_k,1 \leq k \leq r\}$ . The remaining
operators are 
obtained by applying the group $\G$ to each of the generators.

Since the optimal measurement operators satisfy $\Pi_{ik}=U_i \Pi_k U_i^*$
and $\rho_{ik}=U_i \rho_k U_i^*$, $\tr(\rho_{ik} \Pi_{ik})=\tr (\rho_k
\Pi_k)$, so that the 
problem (\ref{eq:pe}) reduces to
the maximization problem
\begin{equation}
\label{eq:max2}
\max_{\Pi_k \in \B} \sum_{k=1}^r\tr(\rho_k\Pi_k),
\end{equation} 
subject to the constraints
\begin{eqnarray}
\label{eq:condp2}
\Pi_k & \geq & 0, \quad 1 \leq k \leq r \nonumber \\
\sum_{i=1}^l\sum_{k=1}^r  U_i \Pi_k U_i^* & = & I.
\end{eqnarray}
Since this problem is a (convex) semidefinite programming problem, the
optimal generators $\Pi_k$ 
can be computed very efficiently in polynomial
time within any desired accuracy \cite{VB96,A91t,NN94}, for example
using the LMI toolbox on Matlab.
Note that the problem of (\ref{eq:max2}) and (\ref{eq:condp2}) has
$rn^2$ real unknowns and 
$r+1$ constraints, in
contrast 
with the original maximization problem  (\ref{eq:pe}) and
(\ref{eq:identu}) which has $lrn^2$ real unknowns
and $lr+1$ constraints.

We summarize our results regarding CGU state sets in the following theorem:
\begin{theorem}[CGU state sets]
\label{thm:cgu}
Let $\SSS = \{\rho_{ik} =
U_i\rho_k U_i^*, U_i \in \G,1 \leq k \leq r\}$
be a compound geometrically uniform (CGU) state set generated
by a finite group $\G$ of unitary matrices and 
generators $\{\rho_k=\phi_k\phi_k^*,1 \leq k \leq r\}$, 
and let $\Phi$ be the matrix of
columns $\phi_{ik}=U_i\phi_k$.  
Then the
least-squares measurement (LSM) is given by the
measurement operators $\Sigma_i=\mu_i\mu_i^*$ with
\[\mu_{ik}=U_i\mu_k\]
where
\[\mu_k=(\Phi\Phi^*)^{-1/2} \phi_k.\]
The LSM has the following properties:
\begin{enumerate}
\item The measurement operators are CGU with generating group $\G$;
\item The probability of correctly detecting each of the states
$\phi_{ik}$ for fixed $k$ using the LSM is the same;
\item If  $\mu_k^*\phi_k=\phi_k^*(\Phi\Phi^*)^{-1/2} \phi_k=\alpha I$
for $1 \leq k \leq r$ then the LSM minimizes the probability of a
detection error.
\end{enumerate}
If in addition the generators $\{\phi_k=V_k\phi,1 \leq k
\leq r\}$ are geometrically uniform
with  $U_iV_k=V_kU_ie^{j\theta(i,k)}$
for all 
$i,k$, then
\begin{enumerate}
\item $\mu_{ik}=U_iV_k\bar{\mu}$ where
$\bar{\mu}=(\Phi\Phi^*)^{-1/2}\phi$ so that the LSM
operators are CGU with geometrically uniform generators;
\item The probability of correctly detecting each of the states
$\phi_{ik}$ using the LSM is the same;
\item If  $\bar{\mu}^*\phi=\phi^*(\Phi\Phi^*)^{-1/2} \phi=\alpha I$
 then the LSM minimizes the probability of a
detection error.
In particular, if $\phi=\ket{\phi}$ is a vector so
that the 
state set is a rank-one 
ensemble, then the LSM minimizes the probability of a detection error.
\end{enumerate}
For arbitrary CGU state sets the optimal measurement operators that
minimize the probability of a detection error are  CGU
with generating group $\G$ and generators $\Pi_k$ that maximize 
$\sum_{k=1}^r \tr(\rho_k \Pi_k)$ subject to  
$\Pi_k \geq 0,1 \leq k \leq r$ and 
$\sum_{i,k} U_i \Pi_k U_i^* = I$.
\end{theorem}

%%%%%%%%%%%%%%%%%%%%%%%%%%%%%%%
\section{Conclusion}
%%%%%%%%%%%%%%%%%%%%%%%%%%%%%%%%

In this paper we considered the optimal measurement operators that
minimize the probability of a detection error when distinguishing
between a collection of {\em mixed} quantum states. We first derived a general
condition under which the LSM minimizes the probability of a detection
error. We then considered state sets with a broad class of symmetry
properties for 
which the LSM is optimal. Specifically, we showed that for GU state
sets and for CGU state sets with generators that satisfy certain
constraints, the LSM is optimal. We also showed that for arbitrary
GU and CGU state sets, the optimal measurement operators have the same
symmetries as the original state sets. Therefore, to compute the
optimal measurement operators, we need only
to compute the corresponding generators. As we showed, the generators can
be computed very efficiently in polynomial time within any desired
accuracy by solving a semidefinite programming problem.

\appendix

%%%%%%%%%%%%%%%%%%%%%%%%%%%%%%%
\section*{\centering Appendix}
%%%%%%%%%%%%%%%%%%%%%%%%%%%%%%%%

%%%%%%%%%%%%%%%%%%%%%%%%%%
\section*{Proof of Theorem~\ref{thm:gcondition}}
%%%%%%%%%%%%%%%%%%%%%%%%%%

In this appendix we prove Theorem~\ref{thm:gcondition}.  Specifically,
we show that for a set of states $\rho_i=\phi_i\phi_i^*$ with prior
probabilities $p_i$, if $\mu_i^*\psi_i=\alpha I $, where
$\mu_i=(\Psi\Psi^*)^{-1/2}\psi_i$ are the LSM factors and
$\psi_i=\sqrt{p_i} \phi_i$, then there exists an Hermitian $X$ such
that
\begin{eqnarray}
\label{eq:appc1}
X &\geq & \psi_i\psi_i^*,\quad 1 \leq i \leq m; \nonumber \\
\label{eq:appc2}
(X-\psi_i\psi_i^*)\mu_i\mu_i^* & = & 0,\quad 1 \leq i \leq m.
\end{eqnarray}

Let $X$ be the symmetric matrix defined by $X=\alpha W^{1/2}$ where
$W=\Psi\Psi^*$. 
Since $\alpha
I=\psi_j^*W^{-1/2}\psi_j=\psi_j^*W^{-1/4}W^{-1/4}\psi_j$, it follows 
that  
\begin{equation}
\label{eq:tmp}
\alpha I \geq W^{-1/4}\psi_j\psi_j^*W^{-1/4}.
\end{equation}
Multiplying both sides of (\ref{eq:tmp}) by $W^{1/4}$ we have
\begin{equation}
\alpha W^{1/2} \geq \psi_j\psi_j^*,
\end{equation}
which verifies that the conditions (\ref{eq:appc1}) are satisfied.  

Next,
\begin{equation}
(X-\psi_i\psi_i^*)\mu_i=
\alpha (\Psi\Psi^*)^{1/2}(\Psi\Psi^*)^{-1/2}\psi_i-\alpha \psi_i=0,
\end{equation}
so that (\ref{eq:appc2}) is also satisfied.

\newpage
\begin{singlespace}

%\bibliography{paper}
%\bibliographystyle{IEEEbib.bst}
\end{singlespace}

\end{document}